\documentclass{article}
\usepackage{amsmath}
\usepackage{amsfonts}
\usepackage{graphicx}
\usepackage{amssymb}
\newtheorem{theorem}{Theorem}

\newtheorem{corollary}[theorem]{Corollary}

\newtheorem{proposition}[theorem]{Proposition}
\newtheorem{remark}[theorem]{Remark}

\newenvironment{proof}[1][Proof]{\noindent\textbf{#1.} }{\ \rule{0.5em}{0.5em}}
\newcommand{\bea}{\begin{align*}}
\newcommand{\eea}{\end{align*}}
\newcommand{\be}{\begin{equation}}
\newcommand{\ee}{\end{equation}}
\newcommand{\ben}{\begin{eqnarray}}
\newcommand{\een}{\end{eqnarray}}
\newcommand{\nd}{\noindent}

\newdimen\dummy
\dummy=\oddsidemargin
\begin{document}

\title{Density operators that extremize Tsallis entropy and thermal stability effects}

\author{C.\ Vignat\\E.E.C.S., University of Michigan, U.S.A. and L.I.S.,
Grenoble, France \\\\ A. Plastino\\ National University La Plata
and Argentina's CONICET\\ C. C. 727, 1900 La Plata, Argentina}

\maketitle

\begin{abstract}
\nd Quite general, analytical (both exact and approximate) forms
for {\it discrete} probability distributions (PD's) that maximize
Tsallis entropy for a fixed variance are here investigated. They
apply, for instance, in  a wide variety of scenarios in which the
system is characterized by a series of discrete eigenstates of the
Hamiltonian. Using these discrete PD's as ``weights" leads to
density operators of a rather general character.  The present
study allows one to vividly exhibit the effects of
non-extensivity. Varying Tsallis'  non-extensivity index $q$ one
is seen to pass from unstable to stable systems and even to
unphysical situations of infinite energy.

\end{abstract}

\section{Introduction}

\nd  Tsallis' thermostatistics is today a new paradigm for
statistics, with applications
 to several scientific disciplines \cite{gellmann,1988,PP1997,FMP,TMP}.
 Notwithstanding  its
 manifold applications, some details of the basic thermostatistical formalism
  remain unexplored. This is why {\it analytical} results are to be
  welcome, specially if they are, as the ones to be here
  investigated, of a very general nature.
  We will provide this type of results for discrete
  probability distributions  of fixed variance that maximize Tsallis' entropy.

\nd Given a discrete probability distribution (DPD)  $p= \{
p_{k} \}$, its Tsallis' information measure (or entropic form) is  defined as
\be
H_{q} (  p )  =\frac{1}{q-1} (  1-\sum_{k=-\infty}^{+\infty}
p_{k}^{q} )  . \label{dino} \ee It is a classical result that
as $q \rightarrow1,$ Tsallis entropy reduces to
Shannon entropy
\[
H_{1} (  p )  =-\sum_{k=-\infty}^{+\infty}p_{k}\log p_{k}.
\]
Without loss of generality, we will here consider only centered
random variables of fixed variance.

\vskip 2mm \nd The aim of this paper is to provide accurate
estimates of i) the parameters of the DPD's and ii) their behavior.
  The maximizers of Tsallis' information measure under variance
constraint in the {\it continuous, multivariate case} have been
discussed in \cite{next2003}. \vskip 3mm \nd  Consider a quantum
system whose eigenstates are characterized by a set of quantum
numbers
 that we collectively denote with an integer, running index $k$ (Cf. Eq.
 (\ref{dino})), that will of course also label the
eigensolutions ($\vert \psi_k \rangle,\,\,\,\,\,\epsilon_k$) of
the  pertinent time-independent Schr\oe dinger equation

\be \label{0q1} \mathcal{H}\, \vert \psi_k \rangle =
\epsilon_k\,\vert \psi_k \rangle,
 \ee with $\mathcal{H}$ the Hamiltonian. Let
 $p_k \equiv \vert \psi_k \vert^2$ be the probability of  finding
 our system in the state $\vert \psi_k \rangle $. The mixed state

 \be \label{0q2} \rho = \sum_k\,p_k\,  \vert \psi_k \rangle \langle \psi_k \vert, \ee
commutes with the Hamiltonian by construction and represents  thus
a bona fide possible stationary state of the system. \nd If we now
find a physical quantity $Z$ whose mean value is proportional to
the variance, we can interpret $\rho$ as the mixed state that
maximizes Tsallis measure subject to the a-priori known
expectation value of such a physical quantity. We will show below
that, in these circumstances,  \fbox{ {\it universal expressions}
can be given for the $p_k$'s, and thus for $\rho$}.

\nd We discuss possible applications in the forthcoming section.
Afterwards, after having introduced some definitions and
notations, we characterize the discrete Tsallis maximizers for
fixed variance in both the $q<1$ and the $q>1$ cases, and discuss
thermal stability questions. We pass then to
 analyze extensions to the multivariate cases.
 For the sake of completeness,  some of the proofs of
assertions  referenced to in what follows are given in  Annex.

\section{Possible physical applications}

\nd Several physical models can be adapted to the scenario
described above (Cf. Eqs. (\ref{0q1}) and (\ref{0q2}))  The
weights $p_k$ in (\ref{0q2}))  will be of the Tsallis-power law
form. With these weights, $\rho$ maximizes Tsallis' entropy
subject to the constraint of a constant variance, which introduces
a Lagrange multiplier that we will call $\beta$.

\nd Given  a physical quantity $Z$ whose mean value is
proportional to the variance,  $\rho$ is that mixed state which
maximizes Tsallis measure subject to the a-priori known $\langle Z
\rangle-$value. In the examples below $Z$ is the system's energy
$E$, but many other possibilities can be imagined. Thus, $\rho$
will be the state maximizing Tsallis' $H_q$ for a fixed value of
the expectation value $U =\langle E\rangle=
Tr\,[\mathcal{H}\,\rho]$ of the Hamiltonian. As a consequence, the
multiplier $\beta$ can be thought of as an ``inverse temperature"
$T$, that is, set $\beta=1/(k_BT)$, with $k_B$ the Boltzmann
constant. This is so because we are at liberty of imagining that
$U$ is kept constant because it is in contact with a heat
reservoir \cite{reif}. Of course, this is not necessarily the
case. $\rho$ exists by itself and is a legitimate stationary mixed
state of our system. But we can think of $\beta$ as either a
``real" or an ``equivalent" inverse temperature.

 \nd Consider, for example,   a system for which the energy spectrum
consists of a denumerable set   of $N$ ($N$ possibly infinite)
 energy levels labeled by a {\it quantum} number $k$ with
 $p_{-k}=p_{k},$
so that all levels exhibit a degeneracy

\be g_k=2\,\,\, for \,\,\,k \ne 0;\,\,\,k > 0;\,\,\,and
\,\,\,g_0=1, \ee i.e., the sums in (\ref{dino}) run from $0$ up to
$\infty$ and each summand is multiplied by $g_k$. Within the
present framework, $U$ \footnote{Mean values linear in the PD are
employed here. They are quite legitimate and were used by Tsallis
in his seminal 1988 paper \cite{1988}. The Legendre structure of
thermodynamics is  definitely respected using them in conjunction
with Tsallis' entropy \cite{PP1997}. Recently, Ferri, Martinez and
Plastino have shown \cite{FMP} that PD's obtained in this manner
can be easily translated into PD's constructed via MaxEnt using
$q-$expectation values evaluated \`a la Tsallis-Mendes-Plastino
\cite{TMP}, showing thereby that there is a one-to-one
correspondence between the two types of PD.}

\be U= \langle E \rangle = \sum_{k=0}^{+\infty} g_k p_{k} E_k,\ee
 becomes numerically equal to the DPD's variance, which by definition is fixed and assumedly known a priori.
 As just stated, we may think,
 if we wish, that our system is in contact with a heat reservoir,
 which fixes the mean energy, and that
 the associated Lagrange multiplier $\beta$
 can be assimilated to an inverse temperature $T$.
 Among many examples of such a scenario we  mention here:

\begin{itemize}
\item
 the planar rotor \cite{dictio}, where $k$ is the magnetic quantum number
corresponding to the azimuthal angle usually denoted by $\phi$.
The level-energies $E_k$ are proportional to  $k^2$ and \be \label{1} E_k = C_E
k^2;\,\,\,C_E\,\,\, {\rm has \,\,\,dimension\,\,\, of\,\,\,
energy.}\ee We have $C_E=\hbar^2/2M_I$, with $M_I$ the system's moment of inertia \cite{Wyllie}.
 For simplicity's sake we take here $C_E=1$, but
retain, of course, its energy-units.
\item the three-dimensional  rigid rotator \cite{dictio}, although $k$ now means the quantum number $L$ associated
to orbital angular momentum (for large $k$, the spectrum of (\ref{1})
looks like that of the 3D rigid rotor) and, for all $k \ge 0$, $g_k=1$.
\item  a vibrating string of length $2l$ with fixed ends
\cite{mandl}, whose energy eigenvalues are $E_k \propto
[k\pi/2l]^2$. \item  the case of a particle (of mass $m$) in a box subject to periodic boundary conditions
(at $(-L,\,\,+L)$) whose energy values are $\epsilon_k=[2\pi^2\hbar^2 k^2/(mL^2)]$ \cite{reif}.
\end{itemize}

\section{Definitions and Notations}

\nd In what follows, $m$ is a positive real number and $n$  a positive
integer. Often, $m$ will take the special form of an odd integer
$m=2n+1.$
The solutions to the problem of maximization of Tsallis
entropy  ``under variance
constraint" (here equivalent to ``for fixed mean energy") will be called discrete Tsallis-probability laws (DTPL's) in
this manuscript. Since (see preceding Section) fixed mean energy can be thought of as implying contact with a heat reservoir at the fixed temperature $T=1/(k_B \beta)$ we will call the pertinent Lagrange multipliers $\beta$ inverse temperatures in what follows.  DTPL's  are given by the following two theorems.

\begin{theorem}
\nd If $\frac{1}{3}<q<1$, for $k_BT=1/\beta$, with $k_B$ the Boltzmann
constant, the discrete probability law $p$ with zero mean and mean
energy $U$ (equivalently, variance $\sigma^{2}$) that maximizes
Tsallis' entropy is defined as\footnote{the dependence of $p$ on
$\beta= 1/k_BT$  is omitted for notational
simplicity}
\begin{equation}
p_{k}=\Pr \{  X=k \}  =f_{q}^{-1} (T  )  (
1+\frac{k^{2}}{k_BT} ) ^{\frac{1}{q-1}}\;\;
\forall
k\in\mathbb{Z^+},
\label{discreteq>1}
\end{equation}
where $f_{q} (  T )  $ is the partition function, given by
\begin{equation}
f_{q} ( T )  =\sum_{k=0}^{+\infty}g_k\, (  1+\frac{k^{2}}
{k_BT} )  ^{\frac{1}{q-1}}, \label{faqle1}
\end{equation}
and $\beta=1/k_BT$ is a real positive Lagrangian multiplier such
that
\[
U \equiv \sigma^{2}=\sum_{k=0}^{+\infty}g_k k^{2}p_{k}.
\]
In the limit $T  \rightarrow \infty$, our probability distribution
converges to the
 classical discrete Gaussian distribution \cite{Zb}.
\end{theorem}

\nd Notice  the existence of a lower  bound $q=1/3$. Smaller $q$
values are unphysical because  for them the mean energy of the
model (for the PD  (\ref{discreteq>1})) becomes infinite.
Additionally (see below) the system turns out to be thermodynamically
unstable because the specific heat becomes negative. In this case where $q<1$, we define
the real positive parameter $m$ that we need below by
\begin{equation}
m=\frac{1+q}{1-q}.
\label{mforq<1}
\end{equation}

\begin{theorem}
\nd If $q>1,$ the discrete probability distribution $p$ with zero mean
and mean energy $U$ (equivalently, variance
$\sigma^{2}$) that maximizes Tsallis entropy is defined as
\ben p_{k}&=&\Pr \{  X=k \}  = \Bigg\{
\begin{array}
[c]{ll}
f_{q}^{-1} ( T )
 ( 1-\frac{k^{2}}{k_BT} ) ^{\frac{1}{q-1}} & \forall
k\in [0, \lfloor k_BT \rfloor -1 ] \\
0 & \text{otherwise}
\end{array}
\label{discreteq<1}
\een
where $f_q  (  T )  $ is the partition function
\begin{equation}
f_{q} ( T )  =\sum_{k=0}^{ \lfloor \beta \rfloor
-1}g_k ( 1-\frac{k^{2}}{k_BT}  ) ^{\frac{1}{q-1}},
\label{faqge1}
\end{equation}
$ \lfloor k_BT \rfloor $ denotes the integer part of
$k_BT$ whose inverse, namely, $\beta$, is a real positive
Lagrangian multiplier such that
\[
U=\sigma^{2}=\sum_{k=0}^{+\infty}g_k\, k^{2}p_{k}.
\]
\end{theorem}
In the present case where $q>1$, we define   $m$ as a real positive parameter related to $q$ by
\begin{equation}
m=\frac{q+1}{q-1},
\label{mforq>1}
\end{equation}
to be of use below. Note that this definition differs from definition (\ref{mforq<1}) in the case $q<1$.

\nd We do not give the proofs of theorems (1) and (2) here. That
laws defined by (\ref{faqle1}) and (\ref{faqge1}) are the
solutions of the Tsallis entropy maximization problem can be
obtained by extending the results of the continuous case as
presented in \cite{next2003}.
 {\it The cases i) $q<1$ and ii) $q>1$ differ essentially by the
fact that the latter has a finite support.} As far as we know,
this  is a novel way, in the $q-$literature, of dealing with the celebrated Tsallis-cut-off \cite{1988}. \vskip 3mm

\nd We remark on the fact that the forms (\ref{discreteq>1}) and
(\ref{discreteq<1})  coincide with the distribution-form given by
Tsallis in his pioneer 1988 paper \cite{1988} (see also
\cite{TMP}), namely, \be \label{tsallis} p_k= Z_q^{-1} [1-(q-1)
\beta^* k^2]^{1/(q-1)},\ee that, for $q  \rightarrow 1$, tends to
the discrete Gaussian  \cite{1988} \be Z_1^{-1}
\,\exp{[-\beta^*k^2]}. \label{gauss}\ee In the former case (Eq.
(\ref{discreteq>1})) we have $q<1$, while in the later (Eq.
(\ref{discreteq<1})) $q>1$. Thus, in Eq. (\ref{discreteq>1}) we
have $\beta= (1-q) \beta^*$, while for (\ref{discreteq<1}) $\beta=
(q-1) \beta^*$. Of course, according to (\ref{discreteq>1}),
(\ref{discreteq<1}), and (\ref{gauss}), for $q \rightarrow 1$ we
have $\beta^* \rightarrow \beta$.
\section{Approximate treatments for $q<1$}

\subsection{The two regimes}

\nd Let us recall that, for $q<1$, $m =(1+q)/(1-q)$. Thus, $m$ is large if
$q \rightarrow 1$ (Boltzmann limit (BL)), while a small positive value of $m$ entails
$q  \rightarrow -1$. Careful inspection of the partition
function $f_{q} ( T ) $ in the cases $m=3
  \leftrightarrow q=1/2$, $m=5 \leftrightarrow q=2/3$,
 and $m=7  \leftrightarrow q=3/4$ indicates that two regimes should be distinguished:

\begin{itemize}
\item the first regime, called \textquotedblright power law\textquotedblright
\ regime, corresponds to the case $T \ll1$ (low temperatures)

\item the second regime called the \textquotedblright
Student-t\textquotedblright\, regime, corresponds to the case $T \gg1$
(high temperatures)
\end{itemize}

\nd Anticipating the results
provided by the following theorems, we note that the cases
\textquotedblright small $T$\textquotedblright/\textquotedblright
large $T$\textquotedblright\ can be translated as joint range
values of parameters $U$ and $m \equiv  (1+q)/(1-q):$
typically, the large $T$ case corresponds to jointly large values
of $U$ and $m,$ as expected,  while the small case corresponds to
jointly small values of these parameters. This is illustrated on
the curve below (Fig. 1), where the large $T$ property is
translated as $T\geq100,$ and the small $T$ property as
$T\leq0.01.$ Note that the left bound $n=1/2$ corresponds to
$m=2n+1=2,$ i.e., $q=1/3$, and thus to an (unphysical) infinite
$U$. For economy's sake we set herefrom

\be \label{defa} a=\sqrt{k_BT}. \ee

\subsection{The power law regime}

In the power-law regime ($T\ll1$), a detailed characterization of
the distribution can be obtained for all real values of $m$, as
shown in the following theorem.

\begin{theorem}
$\label{powerlawregime}$\label{power_law} Assuming that
$T\ll1\,\,\,(a \ll 1)$ and with $m$ defined by (\ref{mforq<1}), the following approximations hold:

\begin{enumerate}
\item for the partition function
\[
f_{q} (  T )  \simeq1+2a^{m+1}\zeta ( m+1 ),
\]

\item for the probability law
\begin{equation}
p_{k}\simeq \bigg\{
\begin{array}
[c]{ll}
\frac{a^{m+1}}{1+2a^{m+1}\zeta (  m+1 )  }k^{-m-1} & \text{if }
k\neq0\\
\frac{1}{1+2a^{m+1}\zeta (  m+1 )  } & \text{if }k=0
\end{array}
,  \label{pk_smalla}
\end{equation}
and corresponds thus to a discrete power-law. It tends to a Gaussian for $q  \rightarrow 1$.

\item for the mean energy
\be \label{mono2} U\simeq2a^{m+1}\zeta (  m-1 ). \ee

\item If, moreover, $m=2n+1$ with $n\in\mathbb{N}$, $n\geq1$, the characteristic
function can be approximated as
\[
\phi_{m} (  u )  \simeq\frac{1}{ (
1+2a^{2n+2}\zeta ( 2n+2 )   )  } (
1+2a^{2n+2}\frac{ (  -1 )  ^{n} ( 2\pi )
^{2n+2}}{ (  2n+2 )  !}B_{2n+2} ( \frac{u}{2\pi
} )   ),
\]
where $B_{n} (  u )  $ denotes the Bernoulli polynomial of order $n.$
\end{enumerate}

\begin{proof}
the proof is given in  Annex 1.
\end{proof}
\end{theorem}
The specific heat $C=dU/dT$ is proportional to $(m+1)$. This is
{\it positive}, and thus the system {\it stable}, for all
$\frac{1}{3}<q<1$. \vskip 3mm \nd Notice that, according to
(\ref{pk_smalla}), only the ground state is populated at $T=0$, as
it should. \vskip 3mm

\nd The probability $p_{k}$ decreases exponentially with $k,$
being thus concentrated around its mean value $EX=0.$ As an
example, if $a=0.1,$ the following table shows, for several values
of $q>\frac{1}{3}$, the
first values of $p_{k}.$
\begin{gather*}
\begin{tabular}
[c]{|c|c|c|c|c|}\hline
& $k=0$ & $k=\pm1$ & $k=\pm2$ & $k=\pm3$\\\hline
$q=\frac{1}{2}$ & $0.999\,78$ & $9.\,\allowbreak997\,8\times10^{-5}$ &
$\allowbreak6.\,\allowbreak248\,6\times10^{-6}$ & $\allowbreak1.\,\allowbreak
234\,3\times10^{-6}$\\\hline
$q=\frac{3}{5}$ & $0.999\,98$ & $9.\,\allowbreak999\,8\times10^{-6}$ &
$\allowbreak3.\,\allowbreak124\,9\times10^{-7}$ & $\allowbreak4.\,\allowbreak
115\,1\times10^{-8}$\\\hline
$q=\frac{2}{3}$ & $\simeq1.\,\allowbreak000\,00$ & $1.\,\allowbreak
000\,00\times10^{-6}$ & $\allowbreak1.\,\allowbreak562\,5\times10^{-8}$ &
$\allowbreak1.\,\allowbreak371\,7\times10^{-9}$\\\hline
$q=\frac{5}{7}$ & $\simeq\allowbreak1.\,\allowbreak000\,00$ & $1.\,\allowbreak
000\,00\times10^{-7}$ & $\allowbreak7.\,\allowbreak812\,5\times10^{-10}$ &
$4.\,\allowbreak572\,5\times10^{-11}$\\\hline
\end{tabular}
\\
\text{Table.1: first values of }p_{k}\text{ for several values of }q  \text{ and } a=0.1
\end{gather*}

\subsection{The Student-t regime}

In the case of the Student-t regime (high temperatures), our main
result writes as follows:

\begin{theorem}
$\label{student-t}\forall\varepsilon>0,$ $\exists a_{0}>0$ and
$\exists
m_{0}>0$ such that if $a \geq a_{0}$ and $2<m\leq m_{0}$ then
\begin{equation}
 \label{inequality}
 \left \vert \sum_{k=0}^{+\infty}g_k\, (
1+\frac{k^{2}}{a^{2}} )
^{-\frac{m+1}{2}}-a\sqrt{\pi}\frac{\Gamma (  \frac{m}{2} )  }
{\Gamma (  \frac{m+1}{2} )  } \right \vert \leq\varepsilon.
\end{equation}

\end{theorem}

\nd The proof of this result is given in  Annex 2. The
values of $a_0$ and $m_{0}$ such that (\ref{inequality}) holds are given in
Fig. 2 for $\epsilon=10^{-9}$. As a direct consequence of this result, we have the
following explicit approximations, valid if, simultaneously, i)
$a\gg1\,\,\,(T\gg1)$ and ii) $q  \rightarrow 1$.

\begin{proposition}
If $a\gg1\ $, $m\gg1$, and $\frac{a^{2}}{m}\gg1$, then the
following approximations hold with high accuracy:

\begin{itemize}
\item for the partition function
\[
f_{q} (  a )  \simeq a\sqrt{\pi}\frac{\Gamma (  \frac{m}
{2} )  }{\Gamma (  \frac{m+1}{2} )  },
\]

\item for the probability law (remember that $U=\sigma^2$)
\begin{equation}
p_{k}\simeq\frac{\Gamma (  \frac{m+1}{2} ) }{\Gamma (
\frac {m}{2} )  \Gamma ( \frac{1}{2} )
\sigma\sqrt{m-2}} ( 1+\frac{k^{2}}{\sigma^{2} (
m-2 )  } ) ^{-\frac{m+1}{2}},
\label{pk_largea}
\end{equation}

\item while for the mean energy
\be \label{mono1} U\simeq\frac{a^{2}}{m-2}. \ee
\end{itemize}
\nd {\rm Note that here}
\be \label{r1} (m-2)\,U= k_B \,T = \beta^{-1}, \ee {\rm a result that will be needed below.}

\begin{proof}
the expression of the partition function results from theorem (\ref{student-t}
). Using a by-product of proof of theorem (\ref{powerlawregime}),
the mean energy $U$ reads
writes
\[
 U=a^{2} (  \frac{f_{m-2}}{f_{m}}-1 ) \simeq
a^{2} (
\frac{\frac{\Gamma (  \frac{m-2}{2} )  }{\Gamma (  \frac{m-1}
{2} )  }}{\frac{\Gamma (  \frac{m}{2} )
}{\Gamma ( \frac{m+1}{2} )  }}-1 )
=\frac{a^{2}}{m-2}.
\]
As a consequence, $a\simeq\sqrt{U(m-2)}$ and the law writes
\[
p_{k}\simeq
\frac{\Gamma (  \frac{m+1}{2} )}{\Gamma ( \frac {m}{2} ) \Gamma (\frac{1}{2}) \sqrt{U(m-2)}}
( 1+\frac{k^2 }{U(m-2)})
^{-\frac{m+1}{2}}
\]
\end{proof}
\end{proposition}
Notice that in this case the specific heat is proportional to
$(m-2)^{-1}$. This is negative, and thus the system unstable,
unless $q > 1/3$. In other words, the system is stable in the
interval $]1/3,1].$ We remark also that, as a consequence of
(\ref{mono2}) and (\ref{mono1}), the mean energy $U$ is an
increasing function of $a$ (and thus of $T$), as it
should.
\begin{remark}
A remarkable result here is that expression (\ref{pk_largea}) is
exactly the sampled version - with the same partition function - of the continuous maximizer of the
Tsallis entropy whose expression is recalled here \cite{TsallisStudent}:

\[
f(x)=
\frac{\Gamma (  \frac{m+1}{2} )}{\Gamma ( \frac {m}{2} ) \Gamma (\frac{1}{2}) \sqrt{U(m-2)}}
( 1+\frac{x^2 }{U(m-2)})
^{-\frac{m+1}{2}} \,\, \forall x \in \mathbb{R}
\]

This entails that, at very high $T$, one can approximate sums over
discrete energy levels by an integral. In order to understand the
situation, remember that the results holds for $m\gg1$, which
entails $q \sim 1$, i.e., the Boltzmann limit. In this limit,
replacing sums by integrals is a commonplace text-book procedure.
\end{remark}

\section{The  $q>1$ instance}

\subsection{A general result and its consequences}

Recall that in the case $q>1$, $m=(q+1)/(q-1)$. The equivalent of theorem (\ref{student-t}) writes as
follows:

\begin{theorem}
\label{Student-r}$\forall\varepsilon>0,$ $\exists a_{0}>0$ and
$\exists
m_{0}>0$ such that if $a\geq a_{0}$ and $m\geq m_{0}$ then
\begin{equation}
\left \vert \sum_{k=- (  a-1 )  }^{ a-1} (  1-\frac{k^{2}}{ a^{2}
} )  ^{\frac{m-1}{2}}-\frac{ a\sqrt{\pi}\Gamma (  \frac{m+1}
{2} )  }{\Gamma (  \frac{m}{2}+1 )  } \right \vert \leq
\varepsilon. \label{approx}
\end{equation}

\begin{proof}
The proof of this result is given in Annex 3. It essentially
follows the steps of the proof of theorem (\ref{student-t}).
\end{proof}
\end{theorem}

\nd  We depict in Fig. 3 the area (north-east), delimited in the plane $ ( a,m )$, that contains
 values of $a$ and $m$ for which approximation (\ref{approx})
holds, with $\varepsilon=10^{-10}$.

\begin{corollary}
For $ a\gg1\,\,\,(T\gg 1)$ the mean energy $U$ verifies
\begin{equation}
U \simeq\frac{ a^{2}}{m+2} \label{varapprox}
\end{equation}

\begin{proof}
Denoting $n=\frac{m-1}{2}$ and $f_{n}(a)$ for $f_{q}(a)$, the normalization constant verifies
the following
recurrence
\begin{align*}
f_{n+1} (  a )   &  =\sum_{k=-a}^{+a} (  1-\frac{k^{2}}{a^{2}
} )  ^{n+1}=\sum_{k=-a}^{+a} (
1-\frac{k^{2}}{a^{2}} )
^{n}-\frac{1}{a^{2}}\sum_{k=-a}^{+a}k^{2} (
1-\frac{k^{2}}{a^{2}} )
^{n}\\
&  =f_{n} (  a )  -\frac{U}{a^{2}}f_{n+1} ( a )
\end{align*}
so that
\[
U =a^{2} (
1-\frac{f_{n+1} (  a ) }{f_{n} (  a ) } ) .
\]
Moreover, using the result of theorem (\ref{Student-r}), we have
\[
f_{n} (  a )  \simeq\frac{\Gamma (
\frac{m}{2}+1 )
}{a\sqrt{\pi}\Gamma (  \frac{m+1}{2} )  }
\]
we obtain after some algebra
\be \label{ener} U
\simeq\frac{a^{2}}{m+2}. \ee

\end{proof}
\end{corollary}
We see from the form of (\ref{ener}) that the specific heat $C$ is
proportional to $1/(m+2)$. Thus, the is system stable for all $q
>1$. Also, since $a^2=k_BT$, we see that the mean energy $U$ is an
increasing function of $T$, as it should.

\subsection{Convergence to a sampled Student-r law}

As a consequence of theorem (\ref{Student-r}), the Tsallis
maximizers can be approximated as follows, for $a$ large enough
($T$ large enough).

\begin{theorem}
If $a\gg1$ then the following approximation holds
\begin{equation}
p_{k}\simeq\frac{\Gamma (  \frac{m}{2}+1 )
}{\sqrt{\pi}\Gamma ( \frac{m+1}{2} ) \sqrt{U}
\sqrt{m+2}} (  1-\frac{k^{2}}{U (  m+2 )  } )  ^{\frac{m-1}{2}}. \label{sampled}
\end{equation}

\begin{proof}
>From (\ref{varapprox}), we have
\[
a^{2}\simeq  (  m+2 )U.
\]
and the result follows from Theorem (\ref{Student-r}).
\end{proof}
\end{theorem}

\nd {\rm Note that here}
\be \label{r2} (m+2)\, U= k_B \,T = \beta^{-1}, \ee {\rm a result that will be needed below.}
\nd We remark the similarity to the corresponding $q<1$ probability
law.

\nd Eq. (\ref{sampled}) corresponds to the sampled version of a
continuous Student-r distribution -which maximizes Tsallis entropy
for fixed mean energy \cite{TsallisStudent} - with the same partition function.

\subsection{Detailed results}

In the case where $m$ is an odd integer, more detailed results
can be obtained concerning the behavior of this probability law.

\begin{theorem}
\label{integercase}For $a\gg1,$ if $m=2n+1$ is an odd integer with
$n\geq2$,
then
\begin{equation}
f_{m} (  a )  =\frac{  2^{2n+1} ( n ! )  ^{2}}{(2n+1)!}a+o (  a^{-2} )  \label{fapprox}
\end{equation}

and $\forall a\geq1,$ and if $m=3$
\[
f_{3} (  a )  =a\frac{4}{3}-\frac{a^{-1}}{3}.
\]
Moreover, the mean  energy $U$ verifies
\begin{equation}
U  =\frac{a^{2}}{m+2}+o ( a^{-1} ).
\label{sigmaapprox}
\end{equation}

\begin{proof}
the proof is given in Annex 4.
\end{proof}
\end{theorem}
This quasi linearity of $f_m (a )  $ in $a$ for high
temperatures is illustrated in Fig. 4, while the quadratic
behavior of $U$ vs.  $a$ is depicted in Fig. 5.

\section{Convergence to the discrete normal distribution}

We show now that, in the case where parameter $m$ grows to infinity, both discrete Tsallis distributions corresponding to either $q<1$ or $q>1$ converge to the discrete normal distribution.

\begin{proposition}
If $a\gg1$ and if the mean energy $U$ is fixed to a constant
value, then, as  $m$ grows to infinity (i.e.
$q \rightarrow 1^{\pm}$), both discrete Tsallis probability distributions
(\ref{discreteq>1}) and (\ref{discreteq<1}) converge to the maximizer of Shannon's discrete entropy,
i.e., the discrete normal distribution.

\begin{proof}
in the case $q<1,$ the result follows by taking $m \rightarrow +\infty$ in
(\ref{pk_largea}) with
\[
\lim_{m \rightarrow +\infty}\frac{\Gamma (  \frac{m+1}{2} )  }
{\Gamma (  \frac{m}{2} )  \Gamma (  \frac{1}{2} )
\sqrt{U(m-2})}=\frac{1}{\sqrt{2U\pi}}
\]
and (Cf. (\ref{r1}))
\[
\lim_{m \rightarrow +\infty} (  1+\frac{k^{2}}{U (
m-2 ) } )  ^{-\frac{m+1}{2}}=\exp (
-\frac{k^{2}}{2U} )  .
\]
In the case $q>1,$ the result follows similarly by taking $m \rightarrow
+\infty$ in (\ref{sampled}) with
\[
\lim_{m \rightarrow +\infty}\frac{\Gamma (  \frac{m}{2}+1 )  }
{\sqrt{\pi}\Gamma (  \frac{m+1}{2} )  \sqrt{m+2}}=\frac{1}{\sqrt
{2\pi}}
\]
and (Cf. (\ref{r2}))
\[
\lim_{m \rightarrow +\infty} (  1-\frac{k^{2}}{U ( m+2 ) } )
^{\frac{m-1}{2}}=\exp ( -\frac{k^{2}}{2U} ).
\]

\end{proof}
\end{proposition}
\nd

This result shows that, for high enough temperatures, and if $m \rightarrow \infty$ (or equivalently $q \rightarrow \pm 1$) the distribution that maximizes Tsallis'
entropy  converges to the Boltzmann one, both for $q<1$ and $q>1$.
 Of course, for fixed $T \gg 1$ and $q \rightarrow +\infty$, we obtain the uniform distribution, whereas for $T \ll 1$ and $q \rightarrow +\infty$ we obtain the deterministic case $p_{k} = \delta_{k}$.

\section{Extension to the multivariate case}

In the last part of this communication
 we give a heuristic discussion of
  the multivariate discrete Tsallis laws with $q<1.$

\begin{theorem}
\label{multivariate}
if $\frac{d}{d+2}<q<1$ then the probability
 $ p_{k_{1},\dots,k_{d}}=Pr\{X_{1}=k_{1},\dots,X_{d}=k_{d}\} $ defined by

\[
p_{k_{1},\dots,k_{d}} =f_{m,d}^{-1} (a_{1},\dots,a_{d} )   (  1+\sum_{i=1}
^{d}\frac{k_{i}^{2}}{a_{i}^{2}} )  ^{-\frac{m+d}{2}};\,\,
\,\,k_{i} \in \mathbb{Z}^{+},
\]
maximizes Tsallis $d$-variate entropy

\[
H_{q}(p)=\frac{1}{q-1} (  1-\sum_{k_{1},\dots,k_{d}}p_{k_{1},\dots,k_{d}}
^{q} )
\]
for fixed mean energy
\[
\sum_{k_{i}}\,g_{k_{i}}k_{i}^{2}\,p_{k_{i}}=u_{i}^{2}.
\]
The partition function $f_{m,d}(a_{1},\dots,a_{d} )$ is defined as follows:
\[
f_{m,d} (a_{1},\dots,a_{d} )  =\sum_{k_{1},\dots,k_{d}} (  1+\sum_{i=1}
^{d}\frac{k_{i}^{2}}{a_{i}^{2}} )  ^{-\frac{m+d}{2}}
\]
In the case where $\forall i,$ $a_{i}\gg1,$ this function can be approximated as
\[
f_{m,d} (a_{1},\dots,a_{d} )\simeq\frac{\pi^{\frac{d}{2}}\Gamma (
\frac {m}{2} ) }{\Gamma (  \frac{m+d}{2} )
}\prod_{i=1}^{d}a_{i}.
\]

\begin{proof}
The proof  can be derived from that given for the continuous case
in \cite{next2003}. It is detailed in Annex 5.
\end{proof}
\end{theorem}

The family of multivariate discrete Tsallis maximizers verify approximately, as in the continuous case, the property of invariance by marginalization, as described in the following theorem.

\begin{theorem}
\label{marginal}

The $d^{\prime}-$variate marginal law ($d'\,\leq d-1$) writes
\begin{equation}
p_{k_{1},\dots,k_{d^{\prime}}}=f_{m,d^{\prime}} ( a_{1},\dots,a_{d^{\prime}} )
 ( 1+\sum_{i=1}^{d^{\prime}}\frac{k_{i}^{2}}{a_{i}^{2}} )
^{-\frac {m+d^{\prime}}{2}}.
\label{renormalizability}
\end{equation}

\begin{proof}
the proof is given in annex 5.
\end{proof}
\end{theorem}
We remark here that property (\ref{renormalizability}) is nothing but
renormalizability: any subsystem of a system distributed
according to a discrete Tsallis law is itself distributed
according to a discrete Tsallis law. We note, too, that renormalization does not change the value of parameter $m$.

\section{Conclusions}

In this report, we have derived {\it in analytic fashion} some elementary properties of the discrete univariate and multivariate Tsallis laws (probability distributions $p_k$
running over a discrete index $k$) for both $q<1$ and $q>1.$
Most of the ensuing results are given by approximations, but
numerical simulations show that they can be regarded as very
accurate ones.

\nd The discrete probability distributions $p_k$ define, for any Hamiltonian $\mathcal{H}$ whose eigenvalue equation reads $\mathcal{H}\vert \psi_k\rangle =\epsilon_k \vert \psi_k\rangle$ mixed states of the form \be \label{qqq2} \rho = \sum_k\,p_k\,  \vert \psi_k \rangle \langle \psi_k \vert, \ee
that commute with the Hamiltonian by construction and represent  thus  bona fide  stationary states of the system defined by $\mathcal{H}$. These weights $p_k$ are of the Tsallis-power law form. With these weights, $\rho$ maximizes Tsallis' entropy subject to variance constraint, which introduces a Lagrange multiplier $\beta$. In this work we have considered situations for which the system's mean energy is proportional to the variance.

\nd Our present consideration allows one to nitidly appreciate the
effects of non-extensivity, as, varying $q$, one passes from
unstable to stable systems and even to unphysical situations of
infinite energy.

\newpage \nd {\bf FIGURE CAPTIONS}

Fig. 1: Plot of $\log_{10}{(\sqrt{U})}$ vs. $[1/(q-1)]$
for $T=100$ (upper curve) and $T=0.01$ (lower curve) that
illustrates the regions of applicability of the Student-t and
power law regimes.

Fig. 2: Student-t regime for $q<1$: the  values of
$a_{0}$ as a function of $m_{0}$ (see text).

Fig. 3: The plane $ ( a,m )$:  values of $a$ and $m$ for
which approximation (\ref{approx}) holds, with
$\varepsilon=10^{-10}$ (see text).

Fig. 4: $q>1$ regime: quasi linearity of $f_m (a ) $ in
$a$ for high temperatures.  The four different curves correspond,
respectively,  to m=3,5, 7, 9, and 11, from top to bottom. Dots
indicate the approximate value as given by Eq. (17). The
continuous line represent  exact values.

Fig. 5:  $q>1$ regime: quadratic behavior of $U$ vs. $a$. The
three curves correspond, respectively,  to  m=3, 5, and 7, from
top to bottom. Dots indicate the approximate value as given by Eq.
(18). The continuous line represent  exact values.

\newpage

\section*{Annex: proofs}
For the sake of completeness we give here some of the proofs of
assertions  referenced to in the text.
\subsection{Annex 1: proof of theorem (\ref{powerlawregime})}
\begin{enumerate}
\item the normalizing factor writes
\begin{align*}
f_{q}(a)   &  =\sum_{k=-\infty}^{+\infty}
(1+\frac{k^{2}}{a^{2}})  ^{-\frac{m+1}{2}}=1+2\sum_{k=1}^{+\infty}
(1+\frac{k^{2}}{a^{2}})  ^{-\frac{m+1}{2}}\\
&  \simeq1+2\sum_{k=1}^{+\infty} (  \frac{k^{2}}{a^{2}} )
^{-\frac{m+1}{2}}=1+2a^{m+1}\zeta(  m+1)
\end{align*}
\item the approximated expression of the distribution as given by
(\ref{pk_smalla}) is a direct consequence of the preceding result

\item We denote indifferently $f_{m}$ as $f_{q}$. As a consequence of the
preceding result and remarking that
\begin{align*}
U  &  =f_{m}^{-1} ( a) a^{2}\sum_{k=-\infty}^{+\infty
}\frac{\frac{k^{2}}{a^{2}}}{ ( 1+\frac{k^{2}}{a^{2}} )  ^{\frac
{m+1}{2}}}\\ &  =f_{m}^{-1} ( a )
a^{2}\sum_{k=-\infty}^{+\infty}\frac {1+\frac{k^{2}}{a^{2}}}{ (
1+\frac{k^{2}}{a^{2}} )  ^{\frac{m+1}{2} }}-f_{m}^{-1} ( a)
a^{2}\sum_{k=-\infty}^{+\infty}\frac{1}{ (
1+\frac{k^{2}}{a^{2}} )  ^{\frac{m+1}{2}}} \\
&  =a^{2} (  \frac{f_{m-2}}{f_{m}}-1 )
\end{align*}
we obtain
\begin{align*}
U  &  =a^{2} (  \frac{f_{2n-1}}{f_{2n+1}}-1 )
\simeq a^{2} (  \frac{1+2a^{2n}\zeta (  2n )
}{1+2a^{2n+2}\zeta (
2n+2 )  }-1 )
=2a^{2n+2} (  \frac{\zeta (  2n )
-a^{2}\zeta (
2n+2 )  }{1+2a^{2n+2}\zeta (  2n+2 )  } ) \\
&  \simeq2a^{2n+2} (  \zeta (  2n )
-a^{2}\zeta (
2n+2 )   )   (  1-2a^{2n+2}\zeta (  2n+2 )   )   \simeq2a^{m+1}\zeta (  m-1 )
\end{align*}
\item if $m=2n+1$ with $n\in\mathbb{N}$, the characteristic function writes
\begin{align*}
\phi_{X} (  u )   &  =f_{m}^{-1} (  a )
\sum_{k=-\infty }^{+\infty}\cos (  ku )   (
1+\frac{k^{2}}{a^{2}} )
^{-\frac{m+1}{2}}
&  =f_{m}^{-1} (  a )   (
1+2\sum_{k=1}^{+\infty}\cos (
ku )   (  1+\frac{k^{2}}{a^{2}} )  ^{-\frac{m+1}{2}} ) \\
&  \simeq f_{m}^{-1} (  a )   (
1+2a^{m+1}\sum_{k=1}^{+\infty }\frac{\cos (  ku )
}{k^{m+1}} )
\end{align*}
But \cite[1.443]{gradshteyn} writes

\be  \sum_{k=1}^{+\infty}\frac{\cos (  k\pi x ) }{k^{2n}} = \frac{
( -1 ) ^{n-1}}{2}  \frac{ ( 2\pi )  ^{2n}}{2n!} B_{2n} (
\frac{x}{2} ) \quad(0\leq x\leq2) \ee so that

\begin{align*}
\phi_{X} (  u )   &  \simeq f_{m}^{-1} (  a )
 ( 1+2a^{m+1}\sum_{k=1}^{+\infty}\frac{\cos (  ku )
}{k^{m+1}} )
\\
&  \simeq\frac{1}{1+2a^{m+1}\zeta (  m+1 )  }  \\ & (  1+2a^{m+1}
\frac{ (  -1 )  ^{n}}{2}\frac{ (  2\pi ) ^{2n+2}}{ ( 2n+2 )
!}B_{2n+2} (  \frac{u}{2\pi} )   ) \quad (0\leq u\leq2\pi)
\end{align*}
\end{enumerate}
\subsection{Annex 2: proof of theorem (\ref{student-t})}
Denote the Gamma function as
\[
\gamma_{\mu} (  v )  =\frac{1}{\Gamma (  \mu )  }
e^{-v}v^{\mu-1},\;v\geq0.
\]
Using the Gamma integral, we have, with $\mu=\frac{m+1}{2},$
\[
\sum_{k=-\infty}^{+\infty} (  1+\frac{k^{2}}{a^{2}} )  ^{-\mu}
=\sum_{k=-\infty}^{+\infty}\int_{0}^{+\infty}e^{-\frac{k^{2}}{a^{2}}v}
\gamma_{\mu} (  v )  dv.
\]
Consider an arbitrary value $a^{2}$ and divide the integral into two parts:
\[
\int_{0}^{+\infty}e^{-\frac{k^{2}}{a^{2}}v}\gamma_{\mu} (
v )
dv=\int_{0}^{a^{2}}e^{-\frac{k^{2}}{a^{2}}v}\gamma_{\mu} (
v )
dv+\int_{a^{2}}^{+\infty}e^{-\frac{k^{2}}{a^{2}}v}\gamma_{\mu} (
v )  dv.
\]
We will show that
\[
I_{1}=\sum_{k=-\infty}^{+\infty}\int_{a^{2}}^{+\infty}e^{-\frac{k^{2}}{a^{2}
}v}\gamma_{\mu} (  v )  dv
\]
can be made arbitrary small provided $a$ is large enough, and for
that choice of $a,$ the absolute value of $I_{2}=\sum_{k=-\infty}^{+\infty}\int_{0}
^{a^{2}}e^{-\frac{k^{2}}{a^{2}}v}\gamma_{\mu} (  v )
dv-a\sqrt{\pi }\frac{\Gamma (  \mu-\frac{1}{2} )
}{\Gamma (  \mu )  }$ can be made arbitrary small by
choosing $\mu$ small enough.

In term $I_{1}$, $v\geq a^{2}$ so that $\exp (  -\frac{k^{2}}{a^{2}
}v )  \leq\exp (  -k^{2} )  $ so that
\[
I_{1}=\sum_{k=-\infty}^{+\infty}\int_{a^{2}}^{+\infty}e^{-\frac{k^{2}}{a^{2}
}v}\gamma_{\mu} (  v )  dv\leq\sum_{k=-\infty}^{+\infty}e^{-k^{2}
}\int_{a^{2}}^{+\infty}\gamma_{\mu} (  v )  dv\leq2\int_{a^{2}
}^{+\infty}\gamma_{\mu} (  v )  dv
\]
since
$\sum_{k=-\infty}^{+\infty}e^{-k^{2}}\simeq\sqrt{\pi}=1.\,\allowbreak
772\,6$. As $\int_{a^{2}}^{+\infty}\gamma_{\mu} (  v )
dv$ is the residue of the converging integral of a positive
function, a proper choice of $a_{0}$ yields
\[
a\geq a_{0} \implies I_{1}\leq\varepsilon/2.
\]
The second term writes
\begin{align*}
I_{2}  &  =\sum_{k=-\infty}^{+\infty}\int_{0}^{a^{2}}e^{-\frac{k^{2}}{a^{2}}
v}\gamma_{\mu} (  v )  dv-a\sqrt{\pi}\frac{\Gamma (
\mu-\frac
{1}{2} )  }{\Gamma (  \mu )  }\\
&  =\int_{0}^{a^{2}}\sum_{k=-\infty}^{+\infty}e^{-\frac{k^{2}}{a^{2}}v}
\gamma_{\mu} (  v )  dv-a\sqrt{\pi}\frac{\Gamma (
\mu-\frac
{1}{2} )  }{\Gamma (  \mu )  }\\
&  = (  \int_{0}^{a^{2}}\sum_{k=-\infty}^{+\infty}e^{-\frac{k^{2}}{a^{2}
}v}\gamma_{\mu} (  v )  dv-\int_{0}^{a^{2}}\frac{a\sqrt{\pi}}
{\sqrt{v}}\gamma_{\mu} (  v )  dv ) \\
&  + (
\int_{0}^{a^{2}}\frac{a\sqrt{\pi}}{\sqrt{v}}\gamma_{\mu} (
v )  dv-a\sqrt{\pi}\frac{\Gamma (  \mu-\frac{1}{2} )  }
{\Gamma (  \mu )  }. )
\end{align*}
For the first difference, we use the fact that, for $a\geq a_{1}$,
\[
 \vert
\sum_{k=-\infty}^{+\infty}e^{-\frac{k^{2}}{a^{2}}v}-\frac
{a\sqrt{\pi}}{\sqrt{v}} \vert \leq\varepsilon
\]
so that
\[
\int_{0}^{a^{2}} (  \sum_{k=-\infty}^{+\infty}e^{-\frac{k^{2}}{a^{2}}
v}-\frac{a\sqrt{\pi}}{\sqrt{v}} )  \gamma_{\mu} (
v ) dv\leq\varepsilon\int_{0}^{a^{2}}\gamma_{\mu} (
v )  dv
\]
This r.h.s. integral can be $\leq\frac{1}{4}$ as soon as $a\geq
a_{1}.$ Choosing $a\geq\max (  a_{1},a_{2} )  $ yields
\[
 \vert \int_{0}^{a^{2}}\sum_{k=-\infty}^{+\infty}e^{-\frac{k^{2}}{a^{2}}
v}\gamma_{\mu} (  v )
dv-\int_{0}^{a^{2}}\frac{a\sqrt{\pi}}{\sqrt {v}}\gamma_{\mu} (
v )  dv \vert \leq\varepsilon/4.
\]
The second difference writes
\begin{align*}
\int_{0}^{a^{2}}\frac{a\sqrt{\pi}}{\sqrt{v}}\gamma_{\mu} (
v ) dv-a\sqrt{\pi}\frac{\Gamma (  \mu-\frac{1}{2} )
}{\Gamma ( \mu )  }  &
=a\sqrt{\pi}\int_{0}^{a^{2}}\frac{1}{\sqrt{v}}\gamma_{\mu
} (  v )  dv-\int_{0}^{+\infty}\frac{a\sqrt{\pi}}{\sqrt{v}}
\gamma_{\mu} (  v )  dv\\
&  =a\sqrt{\pi}\frac{\Gamma (  \mu-\frac{1}{2} )
}{\Gamma ( \mu )  } (
\int_{0}^{a^{2}}\gamma_{\mu-\frac{1}{2}} (  v )
dv-1 )  .
\end{align*}
Now, for $a\geq\max (  a_{0},a_{1} )  $ fixed, function
$f (\mu )  =a\sqrt{\pi}\frac{\Gamma (  \mu-\frac{1}{2} )  }
{\Gamma (  \mu )  } (  \int_{0}^{a^{2}}\gamma_{\mu-\frac{1}{2}
} (  v )  dv-1 )  $ is continuous in  $0$ and such that
\[
\lim_{\mu \rightarrow 0^{+}}f (  \mu )  =0
\]
so that $\exists\mu_{0}$ such that
\[
\mu\leq\mu_{0} \implies \vert f (  \mu )
 \vert \leq\varepsilon/4.
\]
\subsection{Annex 3: proof of theorem (\ref{Student-r})}
Using the complex version of the Gamma integral as given in \cite[3.382.6]
{gradshteyn}, we have
\[
p^{\nu-1}=\frac{\Gamma (  \nu )  }{2\pi}e^{\beta
p}\int_{-\infty }^{+\infty}\frac{e^{ipx}}{ (  \beta+ix )
^{\nu}}dx
\]
if $p>0,$ $\operatorname{Re} (  \beta )  >0$, $\operatorname{Re}
 (  \nu )  >0 $. Choosing $\nu-1=\frac{m-1}{2}$ and $\beta=1,$ we deduce that
\begin{align*}
 (  1-\frac{k^{2}}{a^{2}} )  ^{\frac{m-1}{2}}  &  =\frac
{\Gamma (  \frac{m+1}{2} )  }{2\pi}e^{1-\frac{k^{2}}{a^{2}}}
\int_{-\infty}^{+\infty}\frac{e^{i (  1-\frac{k^{2}}{a^{2}} )  x}
}{ (  1+ix )  ^{\frac{m+1}{2}}}dx \\&  =\frac{\Gamma (
\frac{m+1}{2} ) }{2\pi}\int_{-\infty}^{+\infty }e^{- (  1+ix )
\frac{k^{2}}{a^{2}}}\frac{e^{1+ix}}{ ( 1+ix ) ^{\frac{m+1}{2}}}dx
\end{align*}
so that
\[
\sum_{k=- (  a-1 )  }^{a-1} (
1-\frac{k^{2}}{a^{2}} )
^{\frac{m-1}{2}}=\frac{\Gamma (  \frac{m+1}{2} )  }{2\pi}
\int_{-\infty}^{+\infty}\sum_{k=- (  a-1 )
}^{a-1}e^{- ( 1+ix )
\frac{k^{2}}{a^{2}}}\frac{e^{1+ix}}{ (  1+ix )
^{\frac{m+1}{2}}}dx.
\]
When $a\gg1,$
\[
\sum_{k=- (  a-1 )  }^{a-1}e^{- (  1+ix )  \frac{k^{2}
}{a^{2}}}\simeq\frac{a\sqrt{\pi}}{\sqrt{1+ix}}
\]
so that
\begin{align*}
f_{m} (  a )   &  \simeq\frac{\Gamma (
\frac{m+1}{2} )
}{2\pi}\int_{-\infty}^{+\infty}\frac{a\sqrt{\pi}}{\sqrt{1+ix}}\frac{e^{1+ix}
}{ (  1+ix )  ^{\frac{m+1}{2}}}dx
&  =\frac{\Gamma (  \frac{m+1}{2} )
a\sqrt{\pi}}{2\pi}\int_{-\infty
}^{+\infty}\frac{e^{1+ix}}{ (  1+ix )  ^{\frac{m}{2}+1}}dx\\
&  =\frac{\Gamma (  \frac{m+1}{2} )
a\sqrt{\pi}}{2\pi}\frac{2\pi }{\Gamma (  \frac{m}{2}+1 )
}=\frac{a\sqrt{\pi}\Gamma ( \frac{m+1}{2} )
}{\Gamma (  \frac{m}{2}+1 )  }.
\end{align*}

\subsection{Annex 4: proof of theorem (\ref{integercase})}
Denote $n=\frac{m-1}{2}$ and $f_{n} (  a ) =\sum_{k=- ( a-1 )  }^{a-1} (
1-\frac{k^{2}}{a^{2}} )  ^{n}.$ If $m$ is an odd integer then
$n$ is integer and $\tilde{f} (  n,a )  $ can be
expanded as follows
\begin{align*}
\tilde{f}_{n} (  a )   &  =\sum_{k=0}^{a-1} (  1-\frac{k^{2}
}{a^{2}} )
^{n}=\sum_{k=0}^{a-1}\sum_{p=0}^{n}\binom{n}{p} (
-\frac{k^{2}}{a^{2}} )  ^{p}
&  =\sum_{p=0}^{n}\binom{n}{p} (  -1 )  ^{p}a^{-2p}\sum_{k=0}
^{a-1}k^{2p}.
\end{align*}
But it is well-known \cite[0.121]{gradshteyn} that, if $p\neq0$,
\begin{align*}
\sum_{k=0}^{a-1}k^{2p}  &  =\frac{1}{2p+1}\sum_{l=0}^{2p}\binom{2p+1}{l}
B_{l}a^{2p+1-l}
&
=\frac{1}{2p+1}a^{2p+1}-\frac{1}{2}a^{2p}+\frac{p}{6}a^{2p-1}+o (
a^{2p-2} )
\end{align*}
since $B_{3}=0,$ and if $p=0,\;\sum_{k=0}^{a-1}k^{2p}=a,$ so that
\begin{align*}
\tilde{f}_{n} (  a )  &=\sum_{p=0}^{n}\binom{n}{p} ( -1 )
^{p}a^{-2p}\sum_{k=0}^{a-1}k^{2p} \\&
=a+\sum_{p=1}^{n}\binom{n}{p} (  -1 ) ^{p}a^{-2p} (  \frac
{1}{2p+1}a^{2p+1}-\frac{1}{2}a^{2p}+\frac{p}{6}a^{2p-1}+o (
a^{2p-2} )   ) \\
&  =a (  \sum_{p=0}^{n}\binom{n}{p}\frac{ (  -1 )  ^{p}}
{2p+1} )  -\frac{1}{2}\sum_{p=1}^{n}\binom{n}{p} (
-1 ) ^{p}+\frac{a^{-1}}{6}\sum_{p=1}^{n}\binom{n}{p} (
-1 ) ^{p}p+o (  a^{-2} )
\end{align*}
Moreover, since
\[
\sum_{p=0}^{n}\binom{n}{p}\frac{ (  -1 )
^{p}}{2p+1}=\frac{\sqrt {\pi}}{2}\frac{\Gamma (  n+1 )
}{\Gamma (  \frac{3}{2}+n )
}=\frac{ (  2^{n}n! )  ^{2}}{ (  2n+1 )  !}
\]
by Euler's duplication formula, and as
\[
n>0 \implies \sum_{p=1}^{n}\binom{n}{p} (  -1 )
^{p}= ( 1-1 )  ^{n}-1=-1
\]
and
\[
\sum_{p=1}^{n}\binom{n}{p} (  -1 )
^{p}p=-n\sum_{p=0}^{n}\binom {n-1}{p} (  -1 )
^{p}=-n\delta_{n,1}=
\Big\{
\begin{array}
[c]{c}
-1\text{ if }n=1\\
0\text{ else }
\end{array}
 .
\]
we get finally, for $n\geq2,$
\begin{align*}
\tilde{f}_{n} (  a )   &  =a\frac{ (  2^{n}n! )  ^{2}
}{ (  2n+1 )  !}+\frac{1}{2}+o (  a^{-2} ) \text{ , }
f_{n} (  a )     =a\frac{2 ( 2^{n}n! )  ^{2}}{ (  2n+1 )
!}+o (  a^{-2} )  .
\end{align*}
and in the case $n=1,$
\begin{align*}
\tilde{f}_{1} (  a )   &  =a\frac{2}{3}+\frac{1}{2}-\frac{a^{-1}}
{6}  \text{ , }
f_{1} (  a )    =a\frac{4}{3}-\frac{a^{-1}}{3}
\end{align*}
The partition function verifies the following recurrence
\begin{align*}
f_{n+1} (  a )   &  =\sum_{k=-a}^{+a} (  1-\frac{k^{2}}{a^{2}
} )  ^{n+1}=\sum_{k=-a}^{+a} (
1-\frac{k^{2}}{a^{2}} )
^{n}-\frac{1}{a^{2}}\sum_{k=-a}^{+a}k^{2} (
1-\frac{k^{2}}{a^{2}} )
^{n}\\
&  =f_{n} (  a )  -\frac{U}{a^{2}}f_{n} (  a )
\end{align*}
so that
\[
U =a^{2} (
1-\frac{f_{n+1} (  a ) }{f_{n} (  a )  } )
.
\]
Since $U  \geq0,$ this is an
alternate proof that
$f_{n} (  a )  $ is decreasing in $n.$ Moreover, using (\ref{fapprox}
), we obtain after some algebra
\[
U  =\frac{a^{2}}{2n+3}+o (
a^{-1} )  .
\]
\section{Annex 5: proof of theorems (\ref{multivariate}) and (\ref{marginal})}
We apply the same technique as in the univariate case, and
represent the partition function as a mixture of discrete Gaussian
using the Gamma integral as follows:
\begin{align*}
f_{m,d} ( a_{1},\dots,a_{d})   &  =\sum_{k_{1},\dots,k_{d}} (  1+\sum_{i=1}
^{d}\frac{k_{i}^{2}}{a_{i}^{2}} )  ^{-\frac{m+d}{2}}\\ &
=\sum_{k_{1},\dots,k_{d}}\int_{0}^{+\infty}e^{-\sum_{i=1}^{d}\frac
{k_{i}^{2}}{a_{i}^{2}}}\gamma_{\frac{m+d}{2}} (  v ) dv=\int
_{0}^{+\infty}\sum_{k_{1},\dots,k_{d}}e^{-\sum_{i=1}^{d}\frac{k_{i}^{2}}
{a_{i}^{2}}}\gamma_{\frac{m+d}{2}} (  v )  dv\\
&  =\int_{0}^{+\infty}\prod_{i=1}^{d} (  \sum_{k_{i}}e^{-\frac{k_{i}^{2}
}{a_{i}^{2}}} )  \gamma_{\frac{m+d}{2}} (  v )
dv\simeq\int _{0}^{+\infty} (
\prod_{i=1}^{d}\frac{a_{i}\sqrt{\pi}}{\sqrt{v}} )
\gamma_{\frac{m+d}{2}} (  v )  dv\\
&  =\pi^{\frac{d}{2}} (  \prod_{i=1}^{d}a_{i} )
\int_{0}^{+\infty
}v^{-\frac{d}{2}}\gamma_{\frac{m+d}{2}} (  v )  dv=\pi^{\frac{d}{2}
} (  \prod_{i=1}^{d}a_{i} )  \frac{\Gamma (
\frac{m}{2} ) }{\Gamma (  \frac{m+d}{2} )  }.
\end{align*}
Using the same approach, the $d^{\prime}$-variate marginal law writes
\begin{align*}
p_{k_{1},\dots,k_{d^{\prime}}}  &  =\sum_{k_{d^{\prime}+1},\dots,k_{d}
}p_{k_{1},\dots,k_{d}} =f_{m,d}^{-1} (   a_{1},\dots,a_{d} )
\sum_{k_{d^{\prime}+1},\dots,k_{d}} (
1+\sum_{i=1}^{d}\frac{k_{i}^{2}}{a_{i}^{2}} )  ^{-\frac{m+d}{2}}\\ &
=f_{m,d}^{-1} (   a_{1},\dots,a_{d}  )
\sum_{k_{d^{\prime}+1},\dots,k_{d}} (  (
1+\sum_{i=1}^{d^{\prime}}\frac{k_{i}^{2}}{a_{i}^{2}} )
+\sum_{i=d^{\prime}+1}^{d}\frac{k_{i}^{2}}{a_{i}^{2}} )  ^{-\frac{m+d}
{2}}\\
&  =f_{m,d}^{-1} (  a_{1},\dots,a_{d}   )
(1+\sum_{i=1}^{d^{\prime}}\frac
{k_{i}^{2}}{a_{i}^{2}} )  ^{-\frac{m+d}{2}}\sum_{k_{d^{\prime}+1}
,\dots,k_{d}} (
1+\sum_{i=d^{\prime}+1}^{d}\frac{k_{i}^{2}}{\tilde
{a}_{i}^{2}} )  ^{-\frac{m+d}{2}}
\end{align*}
with
\[
\tilde{a}_{i}^{2}=a_{i}^{2} (  1+\sum_{i=1}^{d^{\prime}}\frac{k_{i}^{2}
}{a_{i}^{2}} )
\]

But, with $m^{\prime}=m+d^{\prime}$,

 \begin{align*} \sum_{k_{d^{\prime}+1} ,\dots,k_{d}} (
1+\sum_{i=d^{\prime}+1}^{d}\frac{k_{i}^{2}}{\tilde {a}_{i}^{2}} )
^{-\frac{m+d}{2}} &\simeq \pi^{\frac{d-d^{\prime}}{2} } (
\prod_{i=d^{\prime}+1}^{d} \tilde{a}_{i} )  \frac{\Gamma (
\frac{m^{\prime}}{2} ) }{\Gamma (
\frac{m^{\prime}+d-d^{\prime}}{2} )  } \\ & =
\pi^{\frac{d-d^{\prime}}{2} } (  \prod_{i=d^{\prime}+1}^{d} a_{i}
)  \frac{\Gamma ( \frac{m^{\prime}}{2} ) }{\Gamma (
\frac{m^{\prime}+d-d^{\prime}}{2} )  } (
1+\sum_{i=1}^{d^{\prime}}\frac{k_{i}^{2} }{a_{i}^{2}}
)^{\frac{d-d^{\prime}}{2}} \end{align*}

so that
\begin{align*}
p_{k_{1},\dots,k_{d^{\prime}}}  & \simeq
\frac{\pi^{\frac{-d}{2}} \Gamma (
\frac{m+d}{2} ) }{(  \prod_{i=1}^{d}a_{i} ) \Gamma (  \frac{m}{2} )  }
\frac{\pi^{\frac{d-d^{\prime}}{2}
} (  \prod_{i=d^{\prime}+1}^{d} a_{i} ) \Gamma (
\frac{m^{\prime}}{2} ) }{\Gamma (  \frac{m^{\prime}+d-d^{\prime}}{2} )  }
(  1+\sum_{i=1}^{d^{\prime}}\frac{k_{i}^{2}
}{a_{i}^{2}} )^{\frac{d-d^{\prime}}{2}} (
1+\sum_{i=1}^{d^{\prime}}\frac{k_{i}^{2}}{
a_{i}^{2}} )  ^{-\frac{m+d}{2}} \\
& = \frac{ \Gamma (
\frac{m+d^{\prime}}{2} ) }{\pi^{\frac{d^{\prime}}{2}} (  \prod_{i=1}^{d^{\prime}}a_{i} ) \Gamma (  \frac{m}{2} )  }
(1+\sum_{i=1}^{d^{\prime}}\frac{k_{i}^{2}}{
a_{i}^{2}} )  ^{-\frac{m+d^{\prime}}{2}} .
\end{align*}

\end{document}